\documentclass[twocolumn,prl,showpacs,superscriptaddress]{revtex4}
\usepackage{amssymb}
\usepackage{amsmath}
\usepackage{graphicx}
\usepackage{subfigure}
\usepackage{epsfig}
\usepackage{amsfonts}

\begin{document}

\title{Decoherence-Based Quantum Zeno Effect in a Cavity-QED System}
\author{D. Z. Xu}
\affiliation{Institute of Theoretical Physics, Chinese Academy of Sciences, Beijing,
100190, China}
\author{Q. Ai}
\affiliation{Institute of Theoretical Physics, Chinese Academy of
Sciences, Beijing, 100190, China}
\author{C. P. Sun}
\email{suncp@itp.ac.cn}
\affiliation{Institute of Theoretical Physics, Chinese Academy of
Sciences, Beijing, 100190, China}

\begin{abstract}
We present a decoherence-based interpretation for the quantum Zeno
effect (QZE) where measurements are dynamically treated as
dispersive couplings of the measured system to the apparatus, rather
than the von Neumann's projections. It is found that the explicit
dependence of the survival probability on the decoherence time
quantitatively distinguishes this dynamic QZE from the usual one
based on projection measurements. By revisiting the cavity-QED
experiment of the QZE [J. Bernu, \textit{et\;al.}, Phys. Rev. Lett,
\textbf{101}, 180402 (2008)], we suggest an alternative scheme to
verify our theoretical consideration that frequent measurements slow
down the increase of photon number inside a microcavity due to the
nondemolition couplings with the atoms in large detuning.
\end{abstract}

\maketitle

\textit{Introduction -- } It usually follows from the von Neumann's
postulate of wave packet collapse (WPC) that the frequent measurements about
whether the system stays in its initial unstable state would inhibit the
transitions to other states \cite{Misra77}. This inhibition phenomenon is
now called the quantum Zeno paradox or the quantum Zeno effect (QZE). Some
experiments, which claimed the verifications of the QZE for various physical
systems \cite{Itano90,Fischer01,Streed06}, seemed to provide clear evidence
supporting the necessariness of the WPC in the logical system of quantum
mechanics. However, many physicists wondered whether the QZE phenomena were
really rooted in the WPC-based measurement (or called the projection
measurement) \cite%
{Frerichs91,Schulman98,Perse,Cook88,Ballentine91,Petrosky90,Pascazio94,Sun95}%
.

In the early days of the discovery of the QZE, Asher Peres demonstrated that
the QZE-like phenomenon could also be explained in terms of the strong
interaction between the observed system and an external agent \cite{Perse}.
When Itano \textit{et\;al.} carried out a QZE experiment based on the
theoretical proposal of Cook \cite{Cook88} and claimed the role of the
projection measurement \cite{Itano90}, some authors argued that no WPC
really happened since the existing experimental data could also be recovered
by unitary dynamic calculations without invoking the WPC \cite%
{Ballentine91,Petrosky90}. Furthermore, a recent experiment in cavity-QED
system for freezing the growth of the photon number in a cavity was
explained in terms of the WPC-based QZE \cite{Bernu08}. It awakened us to
seriously revisit the problem whether this QZE phenomenon depends on the von
Neumann's postulate \cite{Neumann55}, which lies in the core of Copenhagen's
quantum mechanics interpretation (QMI). We expect the similar experiment and
its extension could provide an accessible way to well clarify the physical
distinguishability of different QMIs in accounting for the QZE.

In this Letter, We generally describe the QZE by a unitary evolution
regarding the quantum measurement as a dispersive coupling for decoherence
\cite{Zurek03,Sun93}. With respect to the system's eigenstates being
measured, the decoherence-based quantum measurement is generally formulated
by a diagonal normal operator valued in the apparatus' observable (we call
it the measurement operator) \cite{Zeh03,Sun95,Sun94}. Then we show the
frequent \textquotedblleft bang-bang\textquotedblright\ insertions of such
measurement operators in the original time evolution decohere the system.
These frequent measurements cancel the off-diagonal elements of the system's
density matrix through the destructive interference. Therefore, the
transitions among the eigenstates of the system are inhibited.

This universal proof deals with quantum measurement as a dynamic dephasing
process, rather than an instantaneous collapse. Thus the measurement time is
introduced as a crucial parameter to signature our decoherence-based model
in contrary to the conventional WPC-based one. By re-considering the
cavity-QED experiment \cite{Bernu08} where the periodically driven cavity
field is measured by the nondemolition dispersive couplings to the injected
off-resonant atoms, we calculate the two-dimensional \textquotedblleft phase
diagrams\textquotedblright\ of an alternative experimental scheme with
respect to the measurement time and the \textquotedblleft
bang-bang\textquotedblright\ time interval. Characterizing the dynamic
nature of the QZE, the dependence of the survival probability on the
measurement time explicitly reflect the experimentally testable difference
between two QMIs related to the WPC and dispersive couplings respectively.

\emph{Decoherence-induced quantum Zeno effect -- }Now we develop a general
approach for QZE based on dynamic description of quantum measurement~\cite%
{Pascazio94, Sun94, Sun95}. The dispersive couplings of the measured system $%
S$ to the apparatus $A$ lead to a time evolution of the total system $S$
plus $A$ from the initial state $\left\vert \varphi (0)\right\rangle
=\sum_{j}c_{j}\left\vert s_{j}\right\rangle \otimes \left\vert
a\right\rangle $ to an entangled state $\left\vert \varphi (t)\right\rangle
=M(t)\left\vert \varphi (0)\right\rangle \equiv \sum_{j}c_{j}\left\vert
s_{j}\right\rangle \otimes \left\vert a_{j}\right\rangle $. Here $\left\vert
s_{j}\right\rangle $ $\left( j=1,2,\ldots \right) $ serves as an orthonormal
basis of the Hilbert space $\mathcal{H}_{S}$ of $S$, while $\left\vert
a\right\rangle $ is the initial state of $A$. The unitary measurement
operator $M(t)$ is a diagonal normal matrix with elements $M_{jj}=\exp (-i%
\hat{h}_{j}t)$ for the branch Hamiltonian $\hat{h}_{j}$ being a Hermitian
operator on the Hilbert space $\mathcal{H}_{A}$ of $A$. The final state $%
\left\vert a_{j}\right\rangle \equiv \left\vert a_{j}(t)\right\rangle =\exp
(-i\hat{h}_{j}t)\left\vert a\right\rangle $ of $A$ corresponds to the
system's state $\left\vert s_{j}\right\rangle $. Obviously, $M(t)$ is
capable of defining a nondemolition measurement~\cite{ndm}. An ideal
measurement could well distinguish the apparatus state $\left\vert
a_{j}\right\rangle $ from $\left\vert a_{j^{\prime }}\right\rangle $, i.e., $%
\left\langle a_{j^{\prime }}|a_{j}\right\rangle =\delta _{jj^{\prime }}$. In
this ideal case, the reduced density matrix of the system is depicted by $%
\rho _{s}(t)=\text{Tr}_{A}(\left\vert \varphi (t)\right\rangle \left\langle
\varphi (t)\right\vert )$ with vanishing off-diagonal elements.

$U(t)$ is defined as the unitary evolution operator of $S$ in the absence of
the above \textquotedblleft measurement\textquotedblright. Then we generally
describe the QZE phenomenon by a unitary evolution matrix $%
U_{c}(t)=U_{c}(\tau ,\tau _{m})=[M(\tau _{m})U(\tau )]^{N}$ (see Fig.~\ref%
{fig:sketch}) with a fixed duration $t=N\tau $. Here $\tau$ indicates a
small time interval for which the system evolves freely, and a measurement
with shorter time $\tau _{m}$ is performed at the end of each $U(\tau )$.
Actually, the free evolution co-exists with the measurements through the
whole QZE process, but it could be ignored when measurement is turned on
since the apparatus induces a fast decoherence. An ideal measurement
requires a very short $\tau _{m}$, but a finite $\tau _{m}$ will reflect the
dynamic feature of the realistic measurements. Usually, $U(\tau )$ does not
commutate with $M(\tau _{m})$ so that it can induce the transitions among
states $\left\vert s_{j}\right\rangle $. We re-write $U_{c}(t)$ as a $N$%
-multi-product
\begin{equation}
U_{c}(\tau ,\tau _{m})=\left[ \prod_{n=1}^{N}U_{n}(\tau)\right] M^{N}\text{,}
\label{eq:mproduct}
\end{equation}
where the factors $U_{n}(\tau )=M^{n}U(\tau )M^{-n}$ for$\ M\equiv M(\tau
_{m})$ and $n=1,2,...,N$. For a very short $\tau $ or a very large $N$, it
could be approximated as $U_{n}(\tau )\simeq 1-i\tau M^{n}HM^{-n}\equiv
1-i\tau H_{n}$. If $M$ is not degenerate, we have
\begin{equation}
U_{c}(\tau ,\tau _{m})\simeq \left( 1-itH_{d}-i\frac{t}{N}S\right) M^{N}
\text{,}
\end{equation}%
where $A_{d}$ and $A_{\text{off}}$ denote the diagonal and off-diagonal
parts of matrix $A$, respectively. The summation $S=\sum_{n}(H_{n})_{\text{%
off}}$ is convergent as $N\rightarrow \infty $ or $\tau \rightarrow 0$ for
fixed $t=N\tau $, since $S=\sum_{j\neq j^{\prime }}\Lambda _{jj^{\prime
}}H_{jj^{\prime }}\left\vert s_{j}\right\rangle \langle s_{j^{\prime }}|$,
where
\begin{equation}
\Lambda _{jj^{\prime }}=\frac{\sin (\frac{1}{2}\tau _{m}N\Delta _{jj^{\prime
}})}{\sin (\frac{1}{2}\tau _{m}\Delta _{jj^{\prime }})}e^{-i\tau
_{m}(N+1)\Delta _{jj^{\prime }}/2}
\end{equation}%
for $\Delta _{jj^{\prime }}=\hat{h}_{j}-\hat{h}_{j^{\prime }}$. $\Lambda
_{jj^{\prime }}$ is a finite number when $\Delta _{jj^{\prime }}\neq 0$,
then in the large-$N$ limit, the QZE is achieved as
\begin{equation}
\lim_{N\rightarrow \infty }U_{c}(\tau ,\tau _{m})\rightarrow e^{-iH_{d}t}[1-i%
\mathcal{O}(\frac{t}{N})]M^{N}\text{.}
\end{equation}%
Therefore, the time evolution with very frequent $M$-kicks will keep the
system in its initial state because $U_{c}(\tau ,\tau _{m})$ approaches a
diagonalized unitary matrix $\exp (-iH_{d}t)$.

This argument generally proves the QZE in a dynamic version. Thus the
frequent measurements (for $N\rightarrow \infty $) based on the decoherence
model indeed result in the QZE even though no WPC is used. We remark that
the similar arguments for the QZE have been given by making use of the von
Neumann's quantum ergodic theorem~\cite{Facchi04}.

\emph{Cavity-QED setup for testing decoherence-based quantum Zeno effect -- }
The experiment based on high-$Q$ superconducting cavity has explicitly
demonstrated the increase of the photon number inside the cavity is
suppressed by the continuous measurements~\cite{Bernu08}. In this
experiment, a series of microwave pulses resonant with the cavity are
injected into the cavity, which corresponds to the $U$-process; between
every two adjacent pulses an ensemble of off-resonant atoms are sent into
the cavity to probe the average photon number, playing the part of the $M$%
-process. A single QND probe is actually a dynamic process and changes the
cavity field by a phase factor instead of its photon number. Even we do not
read out the photon number after each probe, the QND coupling of the cavity
field to the off-resonant atom can result in the phase random in the
accumulation of these phase factors thus leads to freezing the photon number
in its initial state. We propose an alternative cavity-QED scheme to verify
this illustration.

\begin{figure}[tbp]
{\includegraphics[bb=35 430 575 650,width=8 cm]{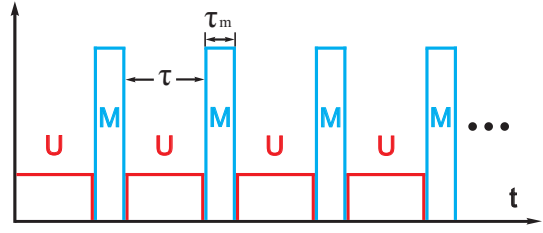}}
\caption{Controlled evolution process containing $N$ unitary evolution $U$
-processes and $N$ dynamic measurement $M$-processes. The $y$-axis
represents the strength of the interaction.}
\label{fig:sketch}
\end{figure}
Let the cavity be initially prepared in the vacuum state $\left\vert
0\right\rangle $ with an ensemble of off-resonant atoms located in it. Then
classical driving laser pulses are sequentially injected into the cavity.
Each pulse is applied for a duration $\tau$. This unitary evolution of the
cavity field is described by the Hamiltonian
\begin{equation}
H_{U}(t)=\omega a^{\dagger }a+fe^{-i\omega _{F}t}a^{\dagger }+h.c.\text{,}
\end{equation}
where $\omega $ is the frequency of the cavity, $f$ and $\omega _{F}$ the
strength and the frequency of the driving field respectively, $a$ and $%
a^{\dagger }$ the annihilation and creation operators of the cavity field.
The driving pulse is peaked at the frequency resonant with the cavity, i.e.,
$\omega _{F}\approx \omega $. Compared to the strength of the driving field,
the interaction between the atom and the cavity field is rather weak, and
thus can be omitted when the pulse is switched on. In the interval when we
turn off the driving field, the atom-field interaction becomes important.
Since the energy level spacing $\omega _{a}$ of the atom and the frequency $%
\omega $ of the cavity are largely detuned, adiabatic elimination results in
an effective measurement Hamiltonian
\begin{equation}
H_{M}=\frac{g^{2}}{\Delta }a^{\dagger }a\left( \left\vert +\right\rangle
\left\langle +\right\vert -\left\vert -\right\rangle \left\langle
-\right\vert \right) \text{.}
\end{equation}%
Here $\left\vert \pm \right\rangle $ are the two atomic energy levels, $g$
the vacuum Rabi frequency defining the atom-cavity coupling and $\Delta
=\omega -\omega _{a}$ the atom-cavity detuning. The unitary evolution
dominated by $H_{M}$ is regarded as a QND measurement, for the atom records
the information of the photon number of the cavity field by its phase of the
$\left\vert \pm \right\rangle $ superposition. The whole experimental
procedure consists of a series of dynamic processes described by $H_{U}$ and
$H_{M}$ alternatively the same as demonstrated in Fig.~\ref{fig:sketch}, but
the strengths of $U$ and $M$ processes are reversed. The probe of the photon
number is only carried out after the last driving pulse.

\emph{Free evolution and decoherence-based measurement -- } The time
evolution of the cavity field governed by $H_{U}\left( \tau \right) $ is
described by phase-modulated displacement operator
\begin{equation}
U(\tau )=e^{i\omega a^{\dagger }a\tau }e^{i\phi \left( \tau \right) }D\left[
\alpha \left( \tau \right) \right] \text{,}
\end{equation}%
where $D\left[ \alpha \left( \tau \right) \right] =\exp [\alpha \left( \tau
\right) a^{\dagger }-\alpha ^{\ast }\left( \tau \right) a]$ with the
displacement parameter $\alpha \left( \tau \right) =\left[ \exp \left(
-i\delta \tau \right) -1\right] f/\delta $, and the phase factor is $\phi
\left( \tau \right) =\left( \sin \delta \tau -\delta \tau \right)
f^{2}/\delta ^{2}$, $\delta =\omega _{F}-\omega $. Here the Wei-Norman
algebra method \cite{Wei63} is used in deriving $U(\tau )$.

In a cavity in the vacuum state $\left\vert 0\right\rangle$, the atom is
initially prepared in the superposition state $\left\vert \phi
(0)\right\rangle =\left(\left\vert +\right\rangle +\left\vert -\right\rangle
\right) /\sqrt{2}$. After the first driving pulse applied for time $\tau $,
the total system evolves into $\left\vert \psi (\tau )\right\rangle
=\left\vert \phi (0)\right\rangle \otimes \left\vert \alpha \left( \tau
\right) \exp \left( -i\omega \tau \right) \right\rangle $. We can see that
the average photon number $\bar{n}=\left\vert \alpha \left( \tau \right)
\right\vert ^{2}\approx f^{2}\tau ^{2}$ quadratically depends on $\tau$, for
$\tau $ is a sufficiently short interval. Then the pulse is turned off and
the atom-cavity field interaction $H_{M}$ dominates the unitary evolution by
$M\left( \tau _{m}\right) =\exp \left( -i\tau _{m}H_{M}\right) $ for the
measurement interval $\tau _{m}$. After the first measurement, the state $%
\left\vert \psi (\tau )\right\rangle $ evolves into an atom-cavity field
entangled state,
\begin{equation}
\left\vert \psi (\tau +\tau _{m})\right\rangle =\frac{1}{\sqrt{2}}e^{i\phi
\left( \tau \right) }\sum_{j=\pm }\left\vert j\right\rangle \otimes
\left\vert \alpha _{j}\right\rangle \text{,}
\end{equation}%
with $\alpha _{\pm }\equiv \alpha \left(\tau \right) \exp \left( -i\omega
\tau \mp ig^{2}\tau _{m}/\Delta \right)$. The average photon number does not
change due to the QND nature of the measurement, but the cavity field
acquires different phases corresponding to the two atomic states.

\emph{Continuous measurement process for QZE -- }During the free evolution,
we insert the decoherence-based measurements for $N$ times at instants $%
n\tau $ $\left( n=1,2,\ldots ,N\right) $. Mathematically, we apply $\left[
M\left( \tau _{m}\right) U\left( \tau \right) \right] ^{N}$ to the initial
state, and then the quantum state evolves into $\left\vert \psi
_{N}\right\rangle \!\!\!\equiv \!\!\!\left\vert \psi \left[ N\left( \tau
+\tau _{m}\right) \right] \right\rangle $=$\sum_{j=\pm }\left[M_{j}\left(
\tau _{m}\right) U\left( \tau \right) \right] ^{N}\left\vert j\right\rangle
\!\otimes \!\left\vert 0\right\rangle $. Here $M\left( \tau _{m}\right) $
acts on the cavity field as two operators $M_{\pm }\left( \tau _{m}\right)
\!\!=\!\!\exp \left(\mp i\xi _{m}a^{\dagger }a\right) $ corresponding to the
two atomic states respectively, where $\xi _{m}\!\!=\!\!g^{2}\tau
_{m}/\Delta $. From the calculations of the explicit expression for $\left[
M_{\pm }\left( \tau _{m}\right) U\left( \tau \right) \right] ^{N}$, we
finally obtain the evolution wavefunction
\begin{equation}
\left\vert \psi _{N}\right\rangle =\sum_{j=\pm }\frac{e^{i\phi _{j}}}{\sqrt{2%
}}\left\vert j\right\rangle \otimes \left\vert \alpha _{jN}e^{-i\omega
t}\right\rangle \text{,}
\end{equation}%
where $\phi _{\pm }=N\phi \left( \tau \right) +\theta _{\pm }\left( N\right)
$, and
\begin{eqnarray*}
\theta _{\pm }\left( N\right) &=&\pm \frac{\left\vert \alpha \left( \tau
\right) \right\vert ^{2}}{2}\frac{N\sin \xi _{m}-\sin \left( N\xi
_{m}\right) }{1-\cos \xi _{m}}\text{,} \\
\alpha _{\pm N} &=&\alpha \left( \tau \right) e^{\mp i\left( N+1\right) \xi
_{m}}\frac{\sin \left( N\xi _{m}/2\right) }{\sin \left( \xi _{m}/2\right)}
\text{.}
\end{eqnarray*}
Accordingly the average photon number is calculated as
\begin{equation}
\bar{n}=\left\vert \alpha \left( \tau \right) \right\vert ^{2}\frac{\sin
^{2}\left( N\xi _{m}/2\right)}{\sin ^{2}\left(\xi _{m}/2\right) }\text{.}
\end{equation}%
We can see in the continuous measurement limit, i.e., $\tau \rightarrow 0$, $%
\left\vert \alpha \left(\tau \right) \right\vert ^{2}\approx f^{2}\tau ^{2}$
.
\begin{figure}[tbp]
{\includegraphics[width=8cm]{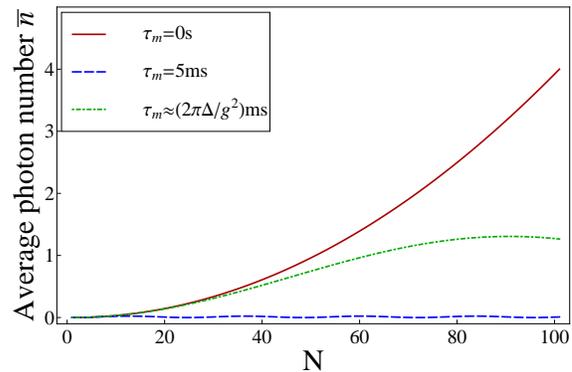}}
\caption{(color online) Average photon number $\bar{n}$ as a function of the
pulse number $N$. We choose $g^{2}/\Delta =10$kHz, $\protect\delta =0.5$Hz, $%
f=400$Hz, and $\protect\tau =50\protect\mu $s. Without the QND probe, $\bar{%
n }$ grows quadratically with $N$ (red solid line). The QZE emerges as $\bar{%
n}$ is frozen at zero with $\protect\tau _{m}=5$ms (bule dashed line). If
the measurement time is chosen specifically at $\protect\tau _{m}=(2\protect%
\pi\Delta /g^{2}+3.5)\protect\mu $s, $\bar{n}$ increases obviously (green
dashdotted line) which is not explained in terms of the WPC interpretation.}
\label{fig:avephotnum}
\end{figure}
Except for certain measurement time interval $\tau_{m}^{\ast}$ chosen as $%
\xi _{m}=g^{2}\tau _{m}^{\ast }/\Delta =2k\pi$, with $k$ integral, $\bar{n}$
approaches zero with $\tau$ decreasing.

As illustrated in Fig.\ref{fig:avephotnum}, $\bar{n}$ shows the similar
inhibition phenomenon (blue dashed line) to Ref. \cite{Bernu08}, with $\tau$
chosen as $50\mu$s. The reason for the photon number ceasing increase is
that the dynamic measurements interrupt the coherent accumulation of photons
by adding a phase factor to the cavity field corresponding to $\xi_{m}$. The
total phase factor after $N$-times measurement destroys the quantum
interference of the cavity field, thus leads to the QZE. This
deocoherence-based process in the existing experiment \cite{Bernu08} reveals
that the QZE can be completely interpreted from the dynamic aspect. To
compare with the situation with only free evolution and no measurements, we
set the atom-cavity coupling $g=0$, and $\bar{n}$ is also depicted in Fig.~%
\ref{fig:avephotnum} (red solid line), which indeed grows quadratically with
$t=N\tau$.

\begin{figure}[tbp]
{\includegraphics[width=7cm]{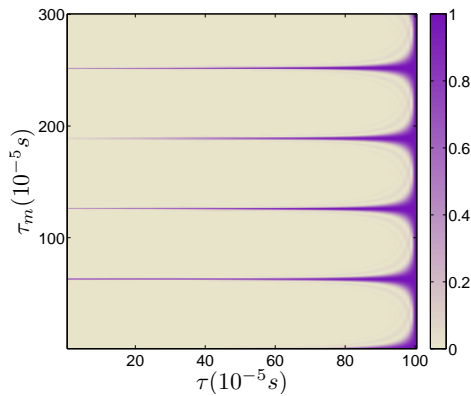}}
\caption{The average photon number as a function of the free evolution time
interval $\protect\tau $ and the measurement time interval $\protect\tau_{m}
$, where $g^{2}/\Delta =10$kHz, $\protect\delta =0.5$Hz, $f=400$Hz, and the
total free evolution time $t$ is fixed at $1$ms. The result is normalized by
the maximum.}
\label{fig:phase}
\end{figure}

The above argument is coincident with the existing experimental data, but
this theoretical description implies the difference between the dynamical
measurement and the projection one. We notice that, when the measurement
time interval is set at critical values $\tau_{m}^{\ast}=2k\pi \Delta /g^{2}$
$(k=1,2,3,...)$, $\bar{n}$ is no longer bounded and increases linearly with $%
N$. In Fig.~\ref{fig:avephotnum}, $\bar{n}$ increases clearly shown as the
green dashdotted line, with $\tau_{m}$ chosen around $\tau _{m}^{\ast }$ as $%
\left(2\pi \Delta /g^{2}+3.5\right) \mu$s. Fixing the total free evolution
time $t$, we illustrate the variation of the average photon number in the
cavity field corresponding to the time interval $\tau$ and $\tau_{m}$ in
Fig.~\ref{fig:phase}. For a given $\tau_{m}$ far from the critical value $%
\tau_{m}^{\ast}$, $\bar{n}$ approaches to zero as $\tau$ decreases, which
recovers the conventional QZE phenomenon based on the projection
measurement. However, $\bar{n}$ mounts up evidently when $\tau_{m}$
approaches to $\tau _{m}^{\ast}$. This $\tau_{m}$-dependent
decoherernce-based QZE could not be predicted by the WPC interpretation,\
but can be testified by the realizable cavity-QED experiment. If we observe
the rise up of the average photon number at certain $\tau_{m}^{\ast}$ in
continuous measurement limit, then we can conclude that the dynamic
measurement model is more compatible with the physical reality in comparison
with the projection measurement in respect of the QZE.

\emph{Conclusion -- }In this Letter, we provided a general algebraic proof
that QZE could be induced by frequent decoherence-based measurements, which
are unitary processes without reference to the WPC postulate (projection
measurement). This approach essentially shows the general QZE phenomenon can
be explained independent of the quantum mechanics-interpretation for the
measurement. Projection measurement provides us a neat description of the
QZE, beyond which, the decoherenced-based model contains more physical
detail. In tht quantum open system, the same model can be extended to
predict the QZE or anti-QZE \cite{Ai10}. Associated with a recent cavity QED experiment
\cite{Bernu08}, we predict an observable effect of the decoherence-based
measurements to distinguish it from the one based on projection measurement:
the survival probability after finite $N$ measurements will explicitly
depend on the measurement time even in the continuous limit. At certain
critical measurement times, the survival probability will deviate from its
initial value predicted in the WPC-based explanation of the QZE.

We thank P. Zhang for valuable discussions. This work was supported by NSFC
through grants 10974209 and 10935010 and by the National 973 program (Grant
No.~2006CB921205).


\smallskip

\begin{thebibliography}{99}
\bibitem{Misra77} B. Misra and E. C. G. Sudarshan, J. Math. Phys. (N.Y.)
\textbf{18}, 756 (1977).

\bibitem{Itano90} W. M. Itano, D. J. Heinzen, J. J. Bollinger and D. J.
Wineland, Phys. Rev. A \textbf{41}, 2295 (1990).

\bibitem{Fischer01} M. C. Fischer, B. Gutierrez-Medina, and M. G. Raizen,
Phys. Rev. Lett. \textbf{87}, 040402 (2001).

\bibitem{Streed06} E. W. Streed, J. Mun, M. Boyd, G. K. Campbell, P. Medley,
W. Ketterle, and D. E. Pritchard, Phys. Rev. Lett. \textbf{97}, 260402
(2006).

\bibitem{Frerichs91} V. Frerichs and A. Schenzle, Phys. Rev. A \textbf{44},
1962 (1991).

\bibitem{Schulman98} L. S. Schulman, Phys. Rev. A \textbf{57}, 1509 (1998).

\bibitem{Perse} A. Peres, Am. J. Phys. \textbf{48}, 931 (1980).

\bibitem{Cook88} R. J. Cook, Phys. Scr. T \textbf{21}, 49 (1988).

\bibitem{Ballentine91} L. E. Ballentine, Phys. Rev. A \textbf{43}, 5165
(1991).

\bibitem{Petrosky90} T. Petrosky, S. Tasaki, and I. Prigogine, Phys. Lett. A
\textbf{151}, 109 (1990); T. Petrosky, S. Tasaki, and I. Prigogine, Physica A
\textbf{170}, 306 (1991).

\bibitem{Pascazio94} S. Pascazio and M. Namiki, Phys. Rev. A \textbf{50},
4582 (1994).

\bibitem{Sun95} C. P. Sun, X. X. Yi and X. J. Liu, Fort. Phys. \textbf{43},
585 (1995).

\bibitem{Bernu08} J. Bernu, S. Del\'{e}glise, C. Sayrin, S. Kuhr, I.
Dotsenko, M. Brune, J. M. Raimond, S. Haroche, Phys. Rev. Lett. \textbf{101}
, 180402 (2008).

\bibitem{Neumann55} J. von Neumann, \textit{Mathematical Foundations of
Quantum Mechanics}, translated by E. T. Beyer (Princeton University Press,
Princeton, 1955).

\bibitem{Zurek03} W. H. Zurek, Rev. Mod. Phys. \textbf{75}, 715 (2003); W.
H. Zurek, Phys. Today. \textbf{44}, 36 (1991).

\bibitem{Sun93} C. P. Sun, Phys. Rev. A \textbf{48}, 898 (1993).

\bibitem{Sun94} C. P. Sun, in \textit{Quantum Classical Correspondence: The
4th Drexel Symposium on Quantum Nonintegrability}, 1994, edited by D. H.
Feng and B. L. Hu (International Press, Cambridge, MA), p. 99.

\bibitem{Zeh03} E. Joos, H. D. Zeh, C. Kiefer, D. J. W. Giulini, J. Kupsch,
I. O. Stamatescu, \textit{Decoherence and the Appearance of a Classical
World in Quantum Theory} (Springer, Berlin, 1996)

\bibitem{ndm} V. B. Braginsky, F. Y. Khalili, K. S. Thorne, \textit{Quantum
Measurement} (Cambridge University Press, Cambridge, UK, 1995)

\bibitem{Facchi04} P. Facchi, D. A. Lidar, and S. Pascazio, Phys. Rev. A
\textbf{69}, 032314 (2004).

\bibitem{Wei63} J. Wei and E. Norman, J. Math. Phys. \textbf{4A}, 575 (1963).

\bibitem{Ai10} Q. Ai, D. Z. Xu, S. Yi, A. G. Kofman, C. P. Sun and F. Nori, in preparation.
\end{thebibliography}
\end{document}